%% file: main.tex
\documentclass[runningheads]{llncs}

\usepackage{graphicx}
\usepackage{amssymb}
\usepackage{multicol}
\usepackage{graphicx}
\usepackage{amsmath}
\usepackage{mathtools}
\usepackage{tikz}
\usepackage{algorithmicx,algorithm}

\usepackage{todonotes}
\newcommand{\kCFC}{{\sc k-Conflict Free Coloring}}

\newcommand{\col}{{\sf col}}
\begin{document}
\title{FPT Algorithms for Conflict-free Coloring of Graphs and Chromatic Terrain Guarding Problems}
\titlerunning{FPT Algorithms for CF-coloring}
%
\author{Akanksha Agrawal\inst{1}\and
Pradeesha Ashok\inst{2} \and
Meghana M Reddy\inst{2} \and
Saket Saurabh\inst{3,4} \and
Dolly Yadav\inst{2}}
\authorrunning{A. Agrawal et al.}
%

\institute{Ben-Gurion University of the Negev, Beer-Sheva, Israel \email{agrawal@post.bgu.ac.il}\\ \and 
International Institute of Information Technology, Bangalore, India \email{pradeesha@iiitb.ac.in , \{meghanam.reddy,dolly.yadav\}@iiitb.org} \and
Institute of Mathematical Sciences, Chennai, India \email{saket@imsc.res.in} \and
University of Bergen, Norway}

%
\maketitle              
\begin{abstract}
\input{abstract.tex}

\end{abstract}

\section{Introduction}
\label{intro}

\input{intro.tex}


\section{Preliminaries}
\label{prelm}

\input{preliminaries.tex}


\section{FPT Algorithm for {\sc{k-Conflict-Free Coloring}}}
\label{res_1}

\input{result1.tex}

\section{FPT Algorithm for {\sc{k-Strong Conflict-Free Coloring}}}
\label{res_2}

\input{result2.tex}


\section{Conflict-Free and Strong Guarding of Terrains}
\label{ter.tex}

\input{terrain.tex}


\bibliography{references}

\newpage
\section*{Appendix}
\input{appendix_2.tex}

\end{document}

%% file: abstract.tex
We present fixed parameter tractable algorithms for the conflict-free coloring problem on graphs. Given a graph $G=(V,E)$, \emph{conflict-free coloring} of $G$ refers to coloring a subset of $V$ such that for every vertex $v$, there is a color that is assigned to exactly one vertex in the closed neighborhood of $v$. The \emph{k-Conflict-free Coloring} problem is to decide whether $G$ can be conflict-free colored using at most $k$ colors. This problem is NP-hard even for $k=1$ and therefore under standard complexity theoretic assumptions, FPT algorithms do not exist when parameterised by the solution size. We consider the \emph{k-Conflict-free Coloring} problem parameterised by the treewidth of the graph and show that this problem is fixed parameter tractable. We also initiate the study of \emph{Strong Conflict-free Coloring} of graphs. Given a graph $G=(V,E)$, \emph{strong conflict-free coloring} of $G$ refers to coloring a subset of $V$ such that every vertex $v$ has at least one colored vertex in its closed neighborhood and moreover all the colored vertices in $v$'s neighborhood have distinct colors. We show that this problem is in FPT when parameterised by both the treewidth and the solution size. We further apply these algorithms to get efficient algorithms for a geometric problem namely the Terrain Guarding problem, when parameterised by a structural parameter.

%% file: intro.tex
Given a graph $G(V,E)$, vertex coloring refers to a function $f:V \rightarrow S$ where $S$ can be considered as a set of colors. Usually various versions of graph coloring impose different restrictions on this coloring function. A popular coloring variant is the \emph{proper coloring} where every vertex should get a color distinct from the colors of all its neighbors.

We consider a graph coloring variant called the \emph{conflict free coloring}.
\begin{definition}
Given a graph $G=(V,E)$, \emph{conflict free coloring} of $G$ refers to coloring a subset of $V$ such that for every vertex $v$, there is a color that is assigned to exactly one vertex in the closed neighborhood of $v$.
\end{definition}

Conflict free coloring was introduced in the context of geometric hypergraphs, motivated by the frequency allocation problem~\cite{smoro,even}. This variant considered coloring a family of geometric objects such that among the objects that have a common intersection, there exists an object of unique color. Pach and Tardos~\cite{pach} studied the problem for graph neighborhoods. All these variants considered coloring all vertices/ objects rather than a subset. However, the number of colors needed in both the variants are asymptotically the same since all the vertices that are not colored can be considered getting a color distinct from all the colored vertices. The variant of conflict free coloring of graphs where only a subset of vertices is colored is studied in~\cite{abel2018,fekete2018}.

Conflict free coloring is also studied in the context of a classic problem in computational geometry, the \emph{Art Gallery} Problem. Given a polygon $P$ with $n$ vertices, $C \subseteq P$ that denotes a set of points in $P$ that need to be guarded and $G \subseteq P$ that denotes a set of points in $P$ where the guards can be placed, the art gallery problem asks to find a subset $G' \subseteq G$ such that every point in $C$ is seen by (guarded by) at least one point in $G'$. Two points see each other if the line segment that connects them lie completely inside $P$.  The \emph{$k$-Conflict-free Art Gallery} problem asks to find a subset $G' \subseteq G$ that can be colored using $k$ colors such that every point in $C$ is seen by a vertex of distinct color in $G'$. This problem has been studied for many classes of polygons~\cite{bartschi2014improved,bartschi2014}. A stronger version of this problem called the \emph{Chromatic Art Gallery} problem is also studied. The \emph{$k$-Chromatic Art Gallery} problem asks to find a subset $G' \subseteq G$ that can be colored using $k$ colors such that every point $p$ in $C$ is seen by at least one vertex in $G'$ and every vertex in $G'$ that sees $p$ is of a different color. This problem was motivated by applications in robotics~\cite{erickson2012}. We generalize this problem to graphs. 
\begin{definition}
Given a graph $G=(V,E)$, \emph{strong conflict-free coloring} of $G$ refers to coloring a subset of $V$ such that for every vertex $v$, there exists at least one colored vertex in the closed neighborhood of $v$ and moreover, every colored vertex in the closed neighborhood of $v$ has a different color.
\end{definition}
Clearly, a strong conflict free coloring is also a conflict free coloring. It is also easy to see that the set of colored vertices in both colorings form a dominating set of the graph. 

\noindent We consider the following algorithmic questions.
\\\noindent {\sc{k-Conflict Free Coloring : }}Given a graph $G=(V,E)$, does there exist a conflict-free coloring of $G$ using at most $k$ colors?
\\\noindent {\sc{k-Strong Conflict Free Coloring : }}Given a graph $G=(V,E)$, does there exist a strong conflict-free coloring of $G$ using at most $k$ colors?

We consider the parameterized complexity of these two questions. These problems are NP-complete even when $k=1$~\cite{abel2018} (Note that both the problems are the same when $k=1$). Hence, they are para-NP hard when parameterized by $k$ and are unlikely to admit FPT algorithms. We study the complexity of the problems when parameterized by the treewidth of the graph. Graphs of bounded treewidth is an important class of graphs that includes outerplanar graphs, Halin graphs and series-parallel graphs. Also many graph problems which are otherwise hard admit FPT algorithms when parameterized by treewidth~\cite{alber2002}.

We further study the parameterized complexity of the chromatic art gallery problem and the conflict free art gallery problem. From a result in \cite{areddy}, $k$-Conflict-free Art Gallery problem and $k$-Chromatic Art Gallery problem are NP-complete for general polygons when $k=1$. Hence the problems are para-NP hard when parameterized by $k$, for general polygons. Here we consider a special class of polygons called \emph{1.5D terrains}. \emph{Terrain Guarding} is a well studied problem in computational geometry and has applications in communication, surveillance and town planning. It is known to be NP-hard~\cite{king2011} and is studied in the area on approximation algorithms~\cite{gibson2009,elbassioni2011}, exact algorithms~\cite{ashok2018} and FPT algorithms~\cite{ashok2018}. Given a terrain $T$ with vertex set $V$, we consider the problem of finding a guard set $V^\prime \subseteq V$ that guards every vertex in $V$. Specifically, we consider the parameterized complexity of conflict free guarding and strong conflict free guarding problems on terrains, when parameterized by a structural parameter called the onion peeling number of the terrains.
\\ \noindent We now give a description of the problems studied in this paper and the results obtained.
\\
\\\noindent \textbf{Problems studied and Results :}

\noindent 1. \textbf{\textsc{k-Conflict free Coloring} problem on graphs} : Given a graph $G(V,E)$ and an integer $k$, does there exist a coloring of a subset of $V$ using at most $k$ colors such that for every vertex $v$, there is a color that is assigned to exactly one vertex in the closed neighborhood of $v$. 
We show that this problem is in FPT when parameterised by the treewidth $\tau$ of the graph $G$.
\\
\\\noindent 2. \textbf{\textsc{k-Strong Conflict free Coloring} problem on graphs} : Given a graph $G(V,E)$ and an integer $k$, does there exist a coloring of a subset of $V$ using at most $k$ colors such that for every vertex $v$, there exists at least one colored vertex in the closed neighborhood of $v$ and moreover, every colored vertex in the closed neighborhood of $v$ has a different color.
 We show that this problem is in FPT when parameterised by $\tau +k$ where $\tau$ is the treewidth of $G$.
 \\
\\\noindent 3.  \textbf{\textsc{Chromatic Terrain Guarding} problem} : Given a terrain $T$ with vertex set $V$, does there exist a coloring on a subset $V'\subseteq V$ such that every vertex $v \in V$ is seen by at least one vertex in $V'$ and moreover all the guards that see $v$ are of different colors.
We show that this problem is in FPT when parameterized by the onion peeling number $p$ of $T$.
\\
\\\noindent 4.  \textbf{\textsc{Conflict free Terrain Guarding} problem} : Given a terrain $T$ with vertex set $V$, does there exist a coloring on a subset $V'\subseteq V$ such that every vertex $v \in V$ is seen by a vertex of distinct color in $V'$.
We show that this problem is in FPT when parameterized by the onion peeling number $p$ of $T$.

%% file: preliminaries.tex
In this section, we discuss some concepts and results that will be used in the subsequent sections.

\noindent
\textbf{Fixed-Parameter Tractability:}
Under standard complexity theoretical assumptions, NP-hard and NP-complete problems are not expected to have polynomial time algorithms. Hence we design algorithms which solve the problem exactly with an exponential running time, but the exponential factor in the running time is restricted to a parameter which is assumed to be small. A problem instance $\pi$, with a parameter $k$, is called \emph{Fixed Parameter Tractable} if there exists an algorithm that solves the problem in time $f(k)\cdot |\pi|^c$, where $c$ is a constant and $f(k)$ is a computable function independent of $|\pi|$. The parameter $k$ is a small positive integer which can be a structural property of either the input or output of $\pi$. The running time of FPT algorithms turns out to be efficient compared to exponential running time algorithms. FPT algorithms and the various techniques can be studied from \cite{PACygan}.




\noindent
\textbf{Treewidth:}
Tree decomposition of a graph $G$ is a pair $(T, X: V(T) \rightarrow 2^{V(G)})$, where $T$ is a tree, and $X_t \subseteq V(G)$ is a vertex subset, where $t$ is a node in the tree $T$. $X_t$ is called the bag of $t$, and the following three conditions hold:
\begin{itemize}
  \item Every vertex of the graph $G$ is in at least one bag.
  \item For every edge $uv \in E(G)$, there is at least one node $t$ of $T$ such that both $u$ and $v$ belong to $X_t$.
  \item For every vertex $v \in V(G)$, the set of nodes of $T$ whose corresponding bags contain $v$, induces a subtree of $T$.
\end{itemize}

The width of a tree decomposition is one less than the maximum size of any bag, i.e., $max_{t \in V(T)}|X_t| - 1$. The $treewidth$ of a graph $G$ is the minimum possible width of a tree decomposition of $G$, and it is denoted by $\tau(G)$. Tree decomposition of a graph is very useful in solving problems. A well-known approach is applying dynamic programming over the tree decomposition of the graph while using the three properties to define the recursion. This technique gives an FPT algorithm for problems like dominating set, vertex cover etc \cite{PACygan}.

\noindent
\textbf{Nice Tree Decomposition:\cite{PACygan,Kloks}}
A tree decomposition with a distinguished root is called a \textit{nice tree decomposition} if:
\\\noindent - All leaf nodes and the root node have empty bags, i.e., $X_l = X_r = \phi$, where $r$ is the root node and $l$ is a leaf node.
  \\\noindent - Every other node in the tree decomposition falls in one of the three categories:
  \\\noindent \textbf{Introduce node:} An introduce vertex node $t$ has exactly one child $t^\prime$ such that $X_t = X_{t^\prime}\cup\{v\}$ for some $v \not\in X_{t^\prime}$.
     \\\noindent \textbf{Forget Node:} A forget node $t$ has exactly one child $t^\prime$ such that $X_t= X_{t^\prime}\setminus \{w\}$ for some $w \in X_t$.
      \\\noindent \textbf{Join Node:} A join node $t$ has exactly two children $t_1$ and $t_2$, such that $X_t= X_{t_1}= X_{t_2}$.
     \\\noindent\textbf{Introduce edge node:} An introduce edge node is labeled with an edge $uv \in E(G)$ such that $u, v \in X_t$ and has exactly one child node $t^\prime$ such that $X_t = X_{t^\prime}$.

Note that every edge of $E(G)$ is introduced exactly once, and we say that the edge $uv$ is \textit{introduced} at $t$. If a join node contains both $u$ and $v$, and the edge $uv$ exists in $E(G)$, we can note that edge $uv$ will be introduced in the subtree above the join node. Nice tree decomposition enables us to add edges and vertices one by one and perform operations accordingly. This variant of tree decomposition still has $O(\tau \cdot n)$ nodes, where $\tau$ is the treewidth of the graph $G$. 

With each node $t$ of the tree decomposition we associate a subgraph $G_t$ of $G$ defined as: $G_t = \left(V_t, E_t = \{e : e \text{ is introduced in the subtree rooted at t\}} \right)$.  Here, $V_t$ is the union of all bags present in the subtree rooted at $t$.

We now state a result regarding computation of a nice-tree decomposition, which follows from~\cite{PACygan,Kloks}. 

\begin{proposition}\label{tw-compute}
Given a graph $G$, in time $O(2^{O(\tau)}n)$, we can compute a nice tree decomposition $(T,{\cal X})$ of $G$ with $|V(T)| \in |V(G)^{O(1)}|$ and of width at most $6\tau$, where $\tau$ is the treewidth of $G$. 
\end{proposition}

\noindent
\textbf{Visibility graphs:} 
Given a polygon $P$, the vertex set of the visibility graph~\cite{ghosh_2007} of $P$ corresponds to the vertex set of $P$ and an edge is added between two vertices in the visibility graph if the corresponding vertices in $P$ see each other. It is easy to observe that a (strong) conflict-free coloring of the visibility graph of $P$ gives us a conflict free (chromatic) guard set in $P$ that guards all the vertices of $P$.

\noindent
\textbf{Terrain Guarding:}
$1.5D$ terrain is an $x-$monotone polygonal chain. An $x$-monotone chain in ${\mathbb{R}^2}$ is a chain that intersects any vertical line at most once. A terrain $T$ consists of a set of vertices $(v_1, v_2, ..., v_n)$, where the x-coordinate of $v_{i+1}$ is greater than the x-coordinate of $v_i$ and there exists an edge $(v_i, v_{i+1})$ for every $i$, $i < n$. Two vertices $v_k$ and $v_l$ are visible to each other if the line segment connecting the two vertices lies entirely above or on the terrain.

\noindent
\textbf{Onion peeling number:}
Onion peeling number of a polygon is the number convex layers of the polygon. For terrains, onion peeling number is defined as the number of upper convex hulls of the terrain. 
\begin{theorem}[\cite{Khodakarami}]\label{thm:khodakarami}
Let $T$ be a terrain with onion peeling number $p$. Then the treewidth of the visibility graph of $T$ is bounded by $2p$. 
\end{theorem}

%% file: result1.tex

In this section, we design an FPT algorithm for the \kCFC\ problem, parameterized by the treewidth of the input graph. The algorithm we design is a dynamic programming over nice tree decomposition. Before moving further, we state a result regarding an upper bound on the number of colors needed to conflict-free color a graph, by the treewidth of it.

\begin{lemma} \label{lemma:treewidth}
The number of colors required to conflict-free color a graph $G$ of treewidth $\tau$ is bounded by $\tau+1$.
\end{lemma}
\begin{proof}
Follows from the facts that a graph $G$ with treewidth $\tau$ is $\tau$-degenerate, i.e., every subgraph has a vertex of degree at most $\tau$ (see for example, Exercise 7.14 in~\cite{PACygan}), $d$-degenerate graphs admit a proper coloring using at most $d+1$ colors (see,~\cite{lick1970k,matula1968min}) and the conflict-free coloring number is upper bounded by the proper coloring number.
\end{proof}

%
 If a vertex $v$ has exactly one vertex, $u\in N[v]$, of color $c_i$ in a coloring, then $v$ is conflict free dominated by $u$ (or $c_i$) in that coloring.
 
The algorithm starts by computing a nice tree decomposition $(T,{\cal X})$ of $G$ in time $O(2^{O(\tau)}n)$, using Proposition~\ref{tw-compute}, of width at most $6\tau$, where $\tau$ is the treewidth of $G$.  

Consider a node $t$ of $T$. We consider a \textit{partitioning} of the bag $X_t$ by a mapping $f : X_t \rightarrow \{B, C, W, R\}$ assigning each vertex in the bag to one of the four partitions. For simplicity, we refer to the vertices in each partition respectively as black, cream, white, and grey vertices. Each vertex is also assigned two more colors by the functions $c : X_t \rightarrow \{c_0, c_1,\ldots, c_k\}$ and $\gamma : X_t \rightarrow \{c_1, c_2,\ldots, c_k\}$. (In the above, $c_0$ will denote a no-color assignment.) Roughly speaking, the colors $c(v)$ and $\gamma(v)$ denote the color assigned to the vertex $v$ and the color which conflict-free dominates $v$ in a conflict free coloring.

\noindent We now give a detailed insight into the partitioning $f$ and colorings $c$ and $\gamma$.
\\\noindent \textbf{Black}, represented by $B$.  A black vertex is assigned the color $c(v)$ in the conflict-free coloring and is conflict-free dominated by color $\gamma(v)$. Note that $c(v) = \gamma(v)$ if the vertex conflict-free dominates itself, and $c(v) \neq c_0$. 
\\\noindent \textbf{Cream}, represented by $C$. Every cream vertex $v$ is given a color (i.e., $c(v) \neq c_0$), but is not conflict-free dominated in the partial solution for $G_t$.  A cream vertex is assigned the color $c(v)$, and is dominated by the color $\gamma(v)$ in the tree above the bag $t$ but not in $G_t$. Note that $c(v) = \gamma(v)$ is not valid for a cream vertex.
\\\noindent \textbf{White}, represented by $W$. A white vertex is not colored in the partial solution for $G_t$, but is conflict-free dominated in the partial solution by a vertex $u$ such that $c(u)=\gamma(v)$. 
\\\noindent \textbf{Grey}, represented by $R$. A vertex $v$ is not colored and is not dominated by the color $\gamma(v)$ in the subgraph $G_t$. (In other words, it will be dominated by a color assigned to a vertex that does not belong to $G_t$.) 


A tuple $(t,c: X_t \rightarrow \{c_0,c_1,\ldots, c_k\}, \gamma: X_t \rightarrow \{c_1,c_2,\ldots, c_k\}, f: X_t \rightarrow \{B,C,W,R\})$ is \emph{valid} if, for each $v \in X_t$, the following holds:
\begin{itemize}
\item if $f(v)=B$, then $c(v) \ne c_0$, 
\item if $f(v)=C$, then $c(v) \ne c_0$ and $c(v) \ne \gamma(v)$,
\item if $f(v)=W$, then $c(v)=c_0$ (and thus, $c(v) \ne \gamma(v)$), and
\item if $f(v)=R$, then $c(v)=c_0$ (and thus, $c(v) \ne \gamma(v)$).
\end{itemize}

For each node $t\in V(T)$ and each valid tuple $(t,c: X_t \rightarrow \{c_0,c_1,\ldots, c_k\}, \gamma: X_t \rightarrow \{c_1,c_2,\ldots, c_k\}, f: X_t \rightarrow \{B,C,W,R\})$, we have a table entry denoted by $d[t, c, \gamma, f]$. We define $d[t, c, \gamma, f] = true$ if and only if $G_t$ admits a coloring ${\sf col}: V_t \rightarrow \{c_0,c_1,\ldots,c_k\}$ (with $c_0$ denoting that no color is assigned to the vertex), such that i) ${\sf col}|_{X_t} = c$,\footnote{For a function $f: X\rightarrow Y$ and a set $X'\subseteq X$, $f|_{X'}$ denotes the function $f$ restricted to the domain $X'$.} ii) for every $v\in X_t$ such that $f(v)\in \{B ,W\}$, there is exactly one vertex $u\in N_{G_{t}}[v]$, with ${\sf col}(u) = \gamma(v)$  iii) for every $v\in  X_t$ such that $f(v)\in \{C , R\}$ and $u\in N_{G_{t}}[v]$, we have ${\sf col}(u) \neq  \gamma(v)$, and  iv) for each $v\in V_t \setminus X_t$, there is a vertex $u\in N_{G_t}[v]$, such that $\col(u) \neq c_0$, and for every other vertex $u' \in N_{G_t}[v]$, $\col(u') \ne \col(u)$.
In the above, such a coloring ${\sf col}$ is called a \emph{$(t, c, \gamma, f)$-good} coloring. (At any point of time wherever we query an invalid tuple, then its value is $false$ by default.) 

Observe that $d[r, \phi, \phi, \phi] = true$ if and only if the graph has a conflict-free coloring using (at most) $k$ colors. (In the above, $\phi$ denotes the function where the domain is the empty set.)

Note that for every node $t$, the set of valid tuples can be found in time bounded by $O(\tau^{O(\tau)})$, as $k$ can be bounded by $\tau+1$ (see Lemma~\ref{lemma:treewidth}). 


We introduce additional notations that will be helpful in stating our algorithm. For a subset $X \subseteq V(G)$, consider a function $f : X \rightarrow \{B,W,G,C\}$. We define $f_{v\rightarrow \alpha} $ where $\alpha \in \{B,W,G,C\}$, as the function where $f_{v\rightarrow \alpha}(x) = f(x)$, if $x\neq v$, and $f_{v\rightarrow \alpha}(x) = \alpha$, otherwise. Similarly, we define $c_{v\rightarrow \alpha}$, for $\alpha \in \{c_0,c_1,\ldots,c_k\}$ and $\gamma_{v\rightarrow \alpha}$ for $\alpha \in \{c_1,c_2,\ldots,$ $c_k\}$.



\noindent We now proceed to define the recursive formulas for computing $d[\cdot]$. \\
\\
\textbf{Leaf node.} For a leaf node $t$, we have $X_t = \emptyset$. Hence, the only (valid) tuple is $d[t, \phi, \phi, \phi]$. We set  $d[t, \phi, \phi, \phi] = true$. The correctness of this step easily follows from the description.\\
\textbf{Introduce vertex node.} Let $t$ be an introduce vertex node with a child $t'$ such that $X_t = X_{t'} \cup \{v\}$ for some $v \not\in X_{t'}$. If $f(v) \in \{C,R\}$, we set $d[t, c, \gamma, f] = d[t', c|_{X'}, \gamma|_{X'}, f|_{X'}]$. If $f(v) =B \text{ and } c(v) = \gamma(v)$, we set $d[t, c, \gamma, f] = d[t', c|_{X'},$ $ \gamma|_{X'}, f|_{X'}]$. Otherwise, we set $d[t, c, \gamma, f] = false$. 

The vertex $v$ is isolated in $G_t$ since no edge incident to $v$ is introduced in this node. Hence in any valid conflict-free coloring of $G_t$, $v$ cannot be conflict-free dominated by any other vertex apart from itself and $v$ cannot conflict-free dominate any other vertex in $G_t$. The correctness of the recurrence formula follows.
\\\textbf{Introduce edge node.} Let $t$ be an introduce edge node labeled with an edge $u^*v^*$ and let $t'$ be the child of it. Thus $G_{t'}$ does not have the edge $u^*v^*$ but $G_t$ has. Consider distinct $u,v \in \{u^*,v^*\}$. We set the value of $d[t, c, \gamma, f]$ based on the following cases. 

\begin{enumerate}
\item If $f(u)=B$, $f(v) = B$, $c(v) = \gamma(u)$, and $c(u) = \gamma(v)$, then we set $d[t,c,\gamma,f] = d[t', c , \gamma ,f_{u\rightarrow C,v\rightarrow C}]$.

\item If $f(u) \in \{B,C\}$, $f(v)= B$, $c(u) = \gamma(v)$, and $c(v) \neq \gamma(u)$, then $d[t, c, \gamma, f] =$ $d[t', c, \gamma,$ $f_{v\rightarrow C}]$.

\item If $f(u) \in \{B,C\}$, $f(v)=W$, and $c(u) = \gamma(v)$, then $d[t, c, \gamma, f] = d[t', c,$ $ \gamma, f_{v\rightarrow R}]$.

\item If $f(u) \in \{B,C\}$, $f(v) \in \{C,R\}$, and $c(u) = \gamma(v)$, then $d[t, c, \gamma, f] = false$.
\item Otherwise, we set $d[t, c, \gamma, f] =  d[t', c, \gamma, f]$.
\end{enumerate}

\begin{lemma}
The recurrence for introduce edge node is correct.
\end{lemma}
\begin{proof}
The proof of forward direction is immediate from the description. We now prove the reverse direction. We prove the correctness for this direction for the case when $f(u)=B$, $f(v) = B$, $c(v) = \gamma(u)$, and $c(u) = \gamma(v)$ (others can be obtained by following similar arguments). Suppose that $d[t', c , \gamma ,f_{u\rightarrow C,v\rightarrow C}] = 1$, and let ${\sf col}: V(G_{t'}) \rightarrow \{c_0,c_1,\ldots, c_k\}$ be a $(t', c , \gamma ,f_{u\rightarrow C,v\rightarrow C})$-good coloring. We show that $\col$ is a $(t, c, \gamma, f)$-good coloring. Notice that in this case, $u^*$ is the unique vertex in $N_{G_t}[v^*]$ with $\col(u^*) = \gamma(v^*)$ (note that $u^*v^* \in E(G_{t})$ and $u^*v^*\notin E(G_{t'})$). Similarly, $v^*$ is the unique vertex in $N_{G_t}[u^*]$ with $\col(v^*) = \gamma(u^*)$. Moreover, for every other vertex $w\in V(G_t)\setminus \{u^*,v^*\} = V(G_{t'})\setminus \{u^*,v^*\}$, $N_{G_{t'}}(w) = N_{G_t}(w)$. Thus we can conclude that $\sf col$ is a $(t,c,\gamma,f)$-good coloring.  

\end{proof}

\noindent\textbf{Forget node.} Let $t$ be a forget node with child $t'$ such that $X_t = X_{t'} \setminus \{w\}$ for some $w \in X_t$. Since the vertex $w$ is not seen again in any node above $t$, the vertex has to be conflict-free dominated in $G_t$, and hence $w$ must be in either the black partition or the white partition. This gives the following recurrence. 
\begin{align*}
    d[t, c, \gamma, f] =  \bigvee_{\substack{i,j \\ 1 \leq i,j \leq k }} \left( d[t', c_{w\rightarrow c_0}, \gamma_{w \rightarrow c_i}, f_{w\rightarrow W}]   \vee  d[t', c_{w\rightarrow c_j}, \gamma_{w \rightarrow c_i}, f_{w\rightarrow B}] \right)
\end{align*}
\\
\textbf{Join node.} Let $t$ be the join node with children $t_1$ and $t_2$. We know that $X_t = X_{t_1} = X_{t_2}$. Recall that in graphs $G_t, G_{t_1}, G_{t_2}$, the set $X_t$ induces an independent set, as the edges are introduced among vertices in $X_t$ after the (topmost) join node. We say that the pair of tuples $(t_1,c_1,\gamma_1,f_1)$ and $(t_2,c_2,\gamma_2,f_2)$  is \textit{$(t,c,\gamma,f)$-consistent} if for every $v \in X_t$, $c(v)=c_1(v)=c_2(v)$ and $\gamma(v)=\gamma_1(v)=\gamma_2(v)$ and one of the following conditions hold:

\begin{itemize}
    \item[1.] $f(v) = B$ and $(f_1(v), f_2(v)) \in \{(B,C), (C,B) \}$ 
    \item[2.] $f(v) = B$ and $(f_1(v), f_2(v)) = (B,B)$ and $c(v) = \gamma(v)$ (here we use the property that $X_t$ is an independent set in the graphs $G_t,G_{t_1}, G_{t_2}$).  
    \item[3.] $f(v) = C$ and $f_1(v)=f_2(v)=C$.
    \item[4.] $f(v) = R$ and $f_1(v)=f_2(v)=R$.
    \item[5.] $f(v)=W$ and $(f_1(v), f_2(v)) \in \{(W,R), (R,W) \}$ (again, here we use the independence property of $X_t$). 
\end{itemize}

We set $d[t, c, \gamma, f] = \bigvee_{(f_1,f_2)} \left(d[t_1, c, \gamma, f_1] \wedge d[t_2, c, \gamma, f_2] \right)$ where $(t_1,c,\gamma,f_1)$ and $(t_2,c,\gamma,f_2)$  is $(t,c,\gamma,f)$-consistent.

We have described the recursive formulas for the values of $d[.]$. Note that we can compute each entry in time bounded by $k^{O(\tau)} n^{O(1)}$. Moreover, the number of (valid) entries for a node $t\in V(T)$ is bounded by $k^{O(\tau)} n^{O(1)}$, and $V(T) \in n^{O(1)}$. Thus we can obtain that the overall running time of the algorithm is bounded by $k^{O(\tau)} n^{O(1)}$. By lemma~\ref{lemma:treewidth}, the following theorem follows.
\begin{theorem}\label{thm:kCFC-FPT}
\kCFC\ is in FPT when parameterized by the treewidth $\tau$.
\end{theorem}


%% file: result2.tex

In this section, we design a fixed parameter tractable algorithm for \textsc{k-Strong Conflict Free Coloring} problem. We obtain our algorithm by doing a dynamic programming over nice tree decomposition.

The algorithm starts by computing a nice tree decomposition $(T,{\cal X})$ of $G$ in time $O(2^{O(\tau)}n)$, using Proposition~\ref{tw-compute}, of width at most $6\tau$, where $\tau$ is the treewidth of $G$.

We define subproblems on $t \in V(T)$ for the graph $G_t$. We consider a \textit{partitioning} of bag $X_t$ by a mapping $f : X_t \rightarrow \{B, W, R\}$.
For simplicity, we refer to the vertices in each partition respectively as black, white and grey. Each vertex is also assigned another color by a function $c$ : $X_{t} \rightarrow \{ c_0,c_1,...,c_k \}$ and a $k$-length tuple, by a function $\Gamma$: $X_{t} \rightarrow \{ 0,1,\hat{1} \}^k$. Roughly speaking, these functions will determine how the ``partial'' conflict-free coloring looks like, when restricted to $G_t$ and vertices of $X_t$. $c(v)$ denotes the color assigned to $v$ and $c(v)=c_0$ denotes that $v$ is not colored. $\Gamma(v)[i]$ indicates whether $v$ has (either in the current graph, or in the ``future'') a vertex in its closed neighborhood that has color $c_i$. $\Gamma(v[i]) = 1$ denotes that vertex $v$ has a vertex in its closed neighborhood of color $c_i$ in $G_t$, $\Gamma(v)[i] = \hat{1}$ denotes that vertex $v$ has a vertex in its closed neighborhood of color $c_i$, that is not present in $G_{t}$, but will appear in the ``future'', and $\Gamma(v)[i] = 0$ denotes the absence of color $c_i$ in the closed neighborhood of $v$. We slightly abuse the notation sometime and use $\Gamma(v)[c_i]$ and $\Gamma(v)[i]$ interchangeably. In the following we give a detailed insight into the functions $f$, $c$ and $\Gamma$.
\\\noindent \textbf{Black}, represented by $B$. Every black vertex $v$ is given a color $c(v) \neq c_0$ in a strong conflict free coloring.
\\\noindent\textbf{Grey}, represented by $R$. A grey vertex $v$ is not colored, i.e. $c(v) = c_0$ and for each $i \in [k]$, it has $\Gamma(v)[i] \in \{0,\hat 1\}$. 
 \\\noindent\textbf{White}, represented by $W$. A vertex $v$ that is neither white nor grey is a white vertex. Note that for a white vertex $v$, $c(v) =c_0$ and there is $i \in [k]$, such that $\Gamma(v)[i]=1$. 
 
We note that although the sets $B,R,W$ are implicit from $\Gamma(\cdot)$, we (redundantly) add them to the tuple for simplicity, and make it more consistent with the previous section. As our goal is to only show whether or not the problem is FPT, we did not try to optimise the running times. 

A tuple $(t,c,\Gamma,f)$ is \emph{valid} if the following conditions hold for every vertex $v \in X_t$: i) $f(v)=B \implies c(v) \ne c_0$ and $\Gamma(v)[c(v)] = 1$, ii) $f(v)=R \implies c(v)=c_0$ and $\Gamma(v)[i] \in \{0,\hat{1}\}$, $\forall i \in \{1...k\}$, and iii) $f(v)=W \implies c(v)=c_0$ and $\Gamma(v)[i] = 1$ for some $i \in \{1...k\}$.

For a node $t\in V(T)$, for each valid tuple $(t,c,\Gamma,f)$, we have a table entry denoted by $D[t,c,\Gamma, f]$. We have $D[t,c,\Gamma, f]= true$ if and only if there is ${\sf col} : V_t \rightarrow \{c_0,c_1,\ldots, c_k\}$ (where $c_0$ denotes no color assignment), such that: \\i) ${\sf col}|_{X_t} = c$, ii) for each $v \in X_t$ and $i\in \{1,2,\dots k\}$ with $\Gamma(v)[i] = 1$, there is exactly one vertex $u\in N_{G_t}[v]$, such that $\col(u) = c_i$, iii) for each $v \in X_t$ and $i\in [k]$ with $\Gamma(v)[i] \in \{0,\hat 1\}$, there is no vertex $u \in N_{G_t}[v]$, such that $\col(u) = c_i$, and iv) for each $v\in V_t \setminus X_t$, there is a vertex $u\in N_{G_t}[v]$, such that $\col(u) \neq c_0$, and for every other $u' \in N_{G_t}[v]$ we have ${\sf col} (u^\prime) \neq {\sf col}(u)$. In the above, such a coloring $\col$ is called a $(t,c,\Gamma,f)$-\emph{good} coloring. (At any point of time wherever we query an invalid tuple, then its value is $false$ by default.) 

 Note that $D[r,\phi,\phi,\phi] = true$, where $r$ is the root of the tree decomposition, if and only if $G$ admits a strong conflict free coloring using (at most) $k$ colors or not. 

\noindent We now proceed to define the recursive formulas for the values of $D$. \\
\noindent\textbf{Leaf node.} For a leaf node $t$, we have $X_t = \emptyset$. Hence, the only entry is $D[t, \phi, \phi, \phi]$. Moreover, by definition, we have $D[t, \phi, \phi, \phi] = true$.\\
\noindent\textbf{Introduce vertex node.} Let $t$ be the introduce vertex node with a child $t'$ such that $X_t = X_{t'} \cup \{v\}$ for some $v \not\in X_{t'}$. Since the vertex $v$ is isolated in $G_t$, the following recurrence follows. If $f(v) = B$ and $\Gamma(v)[i] \in \{0,\hat{1}\}$, for every $i \in \{1,2,\ldots, k\} \setminus \{i^*\}$, where $i^*=c(v)$ (recall that $\Gamma(v)[i^*] = 1$, by the definition of valid tuples, as $f(v) =B$), we set $D[t, c, \Gamma, f] = D[t', c|_{X_{t'}}, \Gamma|_{X_{t'}}, f|_{X_{t'}}]$. If $f(v) = R$, then we set $D[t, c, \Gamma, f] = D[t', c|_{X_{t'}}, \Gamma|_{X_{t'}}, f|_{X_{t'}}]$. Otherwise, we set $D[t, c, \Gamma, f] = false$. 
%
%

\noindent\textbf{Introduce edge node.} Let $t$ be an introduce edge node labeled with an edge $u^*v^*$ and let $t'$ be the child of it. Thus $G_{t'}$ does not have the edge $u^*v^*$ but $G_t$ has. Consider distinct $u,v\in \{u^*,v^*\}$. 
\begin{enumerate}

\item If $f(u)=B$, $f(v)=W$ and $\Gamma(v)[c(u)] = 1$. We set $D[t, c, \Gamma, f]= D[t', c,$ $ \Gamma_{v[c(u)]\rightarrow \hat{1}}$, $f_{v\rightarrow R}] \vee D[t', c, \Gamma_{v[c(u)]\rightarrow \hat{1}}$, $f_{v\rightarrow W}]$ (if any of the entries are invalid, then it is $false$). 


\item If $f(u)=f(v)=B$ and $ \Gamma(v[c(u)]) = \Gamma(u[c(v)]) = 1$, set $D[t, c, \Gamma, f]= D[t', c, \Gamma_{v[c(u)]\rightarrow\hat{1},u[c(v)]\rightarrow\hat{1}}, f] $.
\item If $\{f(u),f(v)\} \cap \{B\} = \emptyset$, then $D[t, c, \Gamma, f]=D[t', c, \Gamma, f]$.
\item If none of the above conditions hold then $D[t, c, \Gamma, f]=false$.
\end{enumerate}
\begin{lemma}
Recurrence for introduce edge node is correct.
\end{lemma}
\begin{proof}
%
%

Note that all the vertices except $u$ and $v$ in $X_t$ are unaffected by introduction of edge $u^*v^*$. Also if neither $u$ nor $v$ are black, then the edge $uv$ cannot strong conflict free dominate any vertex. Therefore the value of $D$ for that tuple is same as that in the child node. 

Consider the case where $f(u) =B$ and $f(v)=W$ and  $\Gamma(v)[c(u)] = 1$. 
We show that $D[t,c,\Gamma,f]=true$ when at least one of $D[t^\prime,c,\Gamma_{v[c(u)]\rightarrow \hat{1}},f_{v\rightarrow R}]$ and $D[t^\prime,c,\Gamma_{v[c(u)]\rightarrow \hat{1}},f]$ is $true$. Let $D[t^\prime,c,\Gamma_{v[c(u)]\rightarrow \hat{1}},f_{v\rightarrow R}]=true$. Let $\col : V_{t\prime} \rightarrow \{c_0,\dots,c_k\}$ be a $(t^\prime,c,\Gamma_{v[c(u)]\rightarrow \hat{1}},f_{v\rightarrow R})$-good coloring in $G_{t^\prime}$. Note that $\col(v) =c_0$ and there is no vertex $v' \in N_{G_{t'}}[v]$ such that $\col(v')=c(u)$. Therefore, $\col$ is also a $(t,c,\Gamma,f)$-good coloring since $u$ is the unique vertex in $N_{G_t}[v]$ with $\col(u)=c(u)$. 
 Hence $D[t,c,\Gamma,f]=true$. Similarly, we can prove that when $D[t^\prime,c,\Gamma_{v[c(u)]\rightarrow \hat{1}},f]=true$, any $(t^\prime,c,\Gamma_{v[c(u)]\rightarrow \hat{1}},f)$-good coloring is also a $(t,c,\Gamma,f)$- coloring. In the reverse direction, assume  $D[t,c,\Gamma,f]=true$ and let $\col : V_{t} \rightarrow \{c_0,\dots,c_k\}$ be a $(t,c,\Gamma,f)$ - good coloring. Therefore, $u$ is the unique vertex in $N_{G_t}[v]$ such that $\col(u)=c(u)$. If there exists a $c_i \ne c(u)$ such that $v$ has a neighbor $v'$ in $N_{G_t}[v]$ with $\col(v')=c_i$, then $\col$ is a $(t^\prime,c,\Gamma_{v[c(u)]\rightarrow \hat{1}},f)$-coloring, otherwise  $\col$ is a $(t^\prime,c,\Gamma_{v[c(u)]\rightarrow \hat{1}},f_{v\rightarrow R})$-coloring.
We can prove the correctness for other cases by similar arguments.



\end{proof}

\textbf{Forget node.} Let $t$ be a forget node with child $t'$ such that $X_t = X_{t'} \setminus \{v\}$ for some $v \in X_{t'}$. Since the vertex $v$ does not appear again in any bag of a node above $t$, $v$ must be either black or white (otherwise, we set the entry to $false$). 
\begin{align*}
    D[t, c, \Gamma, f] =  \bigvee_{\substack{1 \leq i \leq k \\ \alpha \in \{0,1\}^k}} \left( D[t', c_{v\rightarrow c_0}, \Gamma_{v\rightarrow\alpha}, f_{v\rightarrow W}]   \vee  D[t', c_{v\rightarrow c_i}, \Gamma_{v\rightarrow\alpha}, f_{v\rightarrow B}] \right)
\end{align*}
\\
\textbf{Join node.} Let $t$ be the join node with children $t_1$ and $t_2$. We know that $X_t = X_{t_1} = X_{t_2}$ and $X_t$ induces an independent set in the graphs $G_t$, $G_{t_1}$ and $G_{t_2}$. We say that the pair of tuples $[t_1,f_1,c_1,\Gamma_1]$ and  $[t_2,f_2,c_2,\Gamma_2]$ are $[t,f,c,\Gamma]$-consistent if for every $v \in X_t$ the following conditions hold:
\\\noindent -  If $f(v) = B$ then $(f_1(v), f_2(v)) = (B,B)$ and $c_1(v) = c_2(v)=c(v)$. 
   \\\noindent -    If $f(v)=W$ then $(f_1(v), f_2(v)) \in \{(W,R),(R,W),(W,W)\}$.
   \\\noindent -  If $f(v) = R$ then $(f_1(v),f_2(v))=(R,R)$. 
   \\\noindent -  If $\Gamma(v)[i] = 0$ then $(\Gamma_1(v)[i], \Gamma_2(v)[i]) \in \{(0,0)\}$ for $1 \leq i \leq k$.
   \\\noindent -  If $\Gamma(v)[i] = 1$ then $(\Gamma_1(v)[i], \Gamma_2(v)[i]) \in \{(1,\hat{1}),(\hat{1},1)$\} for $1 \leq i \leq k$.
    \\\noindent -   If $\Gamma(v)[i] = \hat{1}$ then $(\Gamma_1(v)[i], \Gamma_2(v)[i]) \in \{(\hat{1},\hat{1})\}$ for $1 \leq i \leq k$.


We set $ D[t, c, \Gamma, f] = \bigvee_{\substack{(f_1,f_2)}}\left({ D[t_1, c_1, \Gamma_1, f_1] \wedge d[t_2, c_2, \Gamma_2, f_2]} \right)$, where $[t_1,f_1$, $c_1,\Gamma_1]$ and $[t_2,f_2,c_2,\Gamma_2]$ is $[t,f,c,\Gamma]$-consistent.
\vspace{5pt}

We have described the recursive formulas for the values of $D[\cdot]$. Note that we can compute each entry in time bounded by $2^{O(k\tau)}\cdot  k^{O(\tau)} n^{O(1)}$. Moreover, the number of (valid) entries for a node $t\in V(T)$ is bounded by $2^{O(k\tau)}\cdot k^{O(\tau)} n^{O(1)}$, and $V(T) \in n^{O(1)}$. Thus we can obtain that the overall running time of the algorithm is bounded by $2^{O(k\tau)} n^{O(1)}$. Thus, we obtain the following theorem.

\begin{theorem}\label{thm:StrongCFC}
\textsc{k-Strong Conflict Free Coloring} is FPT when parameterized by the treewidth $\tau$ of the input graph and the number of colors $k$.
\end{theorem}


%% file: terrain.tex

In this section, we consider the strong chromatic guarding and conflict-free guarding of terrains, parameterized by the onion peeling number of the terrains.

\begin{lemma}\label{thm:terrains}
A $1.5D$ terrain can be strong chromatic guarded using at most $2p$ colors, where $p$ is the onion peeling number of the terrain.
\end{lemma}
We give an algorithm to strong chromatic guard a terrain using $2p$ colors.


Let $T$ be a terrain with vertex set $V$ and let $V$ have $p$ convex layers. Starting with the topmost convex layer i.e layer $1$, we color the guards in each layer using two colors. Let $v$ be the vertex with the highest y-coordinate that lies on the first layer. We color $v$ with color $1$ . The convex layer is now divided into two halves. We go to the right half, and find the first vertex not seen by $v$ and color it with color $2$, and we continue covering the right half this way by alternatively using colors $1$ and $2$. We repeat the same procedure on the left half with the same colors. In the next step, we consider each sub terrain between two consecutive vertices in layer 1 and independently color them in the same way using a different set of colors, say $3$ and $4$. Note that the onion peeling number of any of these sub terrains is at most $p$. In the $i$th layer we have used the colors $2i-1$ and $2i$. Thus we have used at most $2p$ colors. 
%

Clearly every vertex of the terrain is guarded. The vertices that lie in different subterrains in a particular level $j$ do not see each other since there exists at least one vertex from a level $i<j$ that blocks their visibility. Thus it is enough to show that two vertices at the same level are strongly guarded. Two guards $v_1$ and $v_2$ with the same color do not have any overlapping visibility region. By construction there exists a guard $u$ such that $u$ is in the same level between $v_1$ and $v_2$ and has a different color. The guards are chosen such that the visibility region of $v_1$(resp. $v_2$) does not cover the sub terrain between $u$ and $v_1$(resp. $v_2$). Therefore the visibility regions of $v_1$ and $v_2$ do not overlap.
\\

\begin{figure}[t]  
\centering
\begin{tikzpicture}[scale=0.4]  
    
\draw[black, thick] (0.0,0) -- (1.0,3.0); 
\draw[black, thick] (1.0,3.0) -- (1.5,1.0);
\draw[black, thick] (1.5,1.0) -- (2.0,2.0);

\draw[black, thick] (2.0,2.0) -- (3.0,0.5);
\draw[black, thick] (3.0,0.5) -- (3.5,1.5);
\draw[black, thick] (3.5,1.5) -- (4.0,1.0);

\draw[black, thick] (4.0,1.0) -- (4.5,3.0);
\draw[black, thick] (4.5,3.0) -- (5.0,2.5);
\draw[black, thick] (5.0,2.5) -- (6.5,6.0);

\draw[black, thick] (6.5,6.0) --(7.0,2.0);
\draw[black, thick] (7.0,2.0) --(8.0,6.25);
\draw[black, thick] (8.0,6.25) --(9.0,2.0);

\draw[black, thick] (9.0,2.0) -- (10.0,4.0);
\draw[black, thick] (10.0,4.0) -- (10.5,3.0);
\draw[black, thick] (10.5,3.0) -- (11.0,3.5);

\draw[black, thick] (11.0,3.5) -- (12.5,1.5);
\draw[black, thick] (12.5,1.5) -- (13.0,4.5);
\draw[black, thick] (13.0,4.5) -- (14.5,1.0);

\draw[black, thick] (14.5,1.0) -- (15.5,2.5);
\draw[black, thick] (15.5,2.5) -- (16.5,1.0);

\draw[blue, thick, dash dot] (0.0,0) -- (1.0,3.0);
\draw[blue, thick, dash dot] (1.0,3.0) -- (6.5,6.0);
\draw[blue, thick, dash dot] (6.5,6.0) -- (8.0,6.25);
\draw[blue, thick, dash dot] (8.0,6.25) -- (13.0,4.5);
\draw[blue, thick, dash dot] (13.0,4.5)--(15.5,2.5);
\draw[blue, thick, dash dot] (15.5,2.5) -- (16.5,1.0);

\filldraw[blue] (8.0,6.25) circle (2pt) node[anchor=south] {$1$};
\filldraw[blue] (1.0,3.0) circle (2pt) node[anchor=south] {$2$};
\filldraw[blue] (15.5,2.5) circle (2pt) node[anchor=south] {$2$};

\draw[red, thick, dash dot] (1.5,1.0) -- (2.0,2.0);
\draw[red, thick, dash dot] (2.0,2.0) -- (4.5,3.0);
\draw[red, thick, dash dot] (4.5,3.0) -- (5.0,2.5);

\draw[red, thick, dash dot] (9.0,2.0) -- (10.0,4.0);
\draw[red, thick, dash dot] (10.0,4.0)--(11.0,3.5);
\draw[red, thick, dash dot] (11.0,3.5) -- (12.5,1.5);

\filldraw[red] (1.5,1.0) circle (2pt) node[anchor=north] {$4$};
\filldraw[red] (4.5,3.0) circle (2pt) node[anchor=south] {$3$};

\filldraw[red] (10.0,4.0) circle (2pt) node[anchor=south] {$3$};
\filldraw[red] (12.5,1.5) circle (2pt) node[anchor=north] {$4$};

\filldraw[red] (7.0,2.0) circle (2pt) node[anchor=north] {$3$};
\filldraw[red] (14.5,1.0) circle (2pt) node[anchor=north] {$3$};

\end{tikzpicture}%
    \caption{Two convex layers with colored guards} \label{fig:convex_layers}
    
\end{figure}
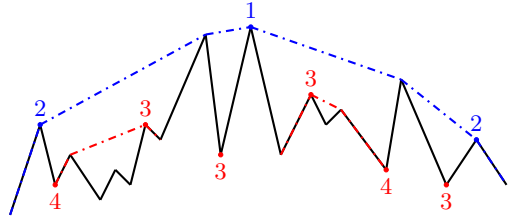

%
%
%
%

The following theorem follows from Theorem \ref{thm:kCFC-FPT} and Lemma \ref{thm:terrains}.
\begin{theorem}
{\sc{Strong Chromatic Guarding}} problem for terrains is in FPT when parameterized by the onion-peeling number of the terrain.
\end{theorem}

By a similar method, we can obtain FPT algorithms for the {\sc{Conflict free Guarding}} problem for terrains, when parameterized by the onion-peeling number.
\begin{lemma}\label{cfcolorslemma}
A $1.5D$ terrain can be conflict-free chromatic guarded using at most $p+1$ colors, where $p$ is the onion peeling number of the terrain. 
\end{lemma}
The proof follows from an algorithm similar to that given in the proof of lemma~\ref{thm:terrains}. Here, we use colors $i$ and $i+1$ to color the vertices in the $i$th convex layer. Hence, only $p+1$ colors are required.

The following theorem follows from Theorem~\ref{thm:StrongCFC} and Lemma~\ref{cfcolorslemma}.
\begin{theorem}
{\sc{Conflict Free Guarding}} problem for terrains is in FPT when parameterized by the onion-peeling number of the terrain.
\end{theorem}

%% file: appendix_2.tex
\begin{lemma}~\label{jointnodelemma1}
Recurrence for the join node for \textsc{k - Conflict free Coloring} is correct.
\end{lemma}
\begin{proof}

Consider the join node $X_t$. Assume $d[t,c,\gamma,f] = true$ and ${\sf col}: V_{t} \rightarrow \{c_0,c_1,\ldots, c_k\}$ be a $(t,c,\gamma,f)$-good coloring. We prove that there exists $[t_1,c_1,\gamma_1,f_1]$ and $[t_2,c_2,\gamma_2,f_2]$, such that the pair is  $(t,c,\gamma,f)$-consistent and $d[t_1,c_1,\gamma_1,f_1] =true$ and $d[t_2,c_2,\gamma_2,f_2] = true$.  For every vertex $v$, we assign $c_1(v)=c_2(v)=c(v)$ and $\gamma_1(v)=\gamma_2(v)=\gamma(v)$. We define $f_1$ and $f_2$ for the vertex $v$ as follows:
\vspace{5pt}

\noindent \emph{Case 1: $f(v)=W$} : Let $\gamma(v)=c_i$.  Let $u$ be the unique neighbor of $v$ in $G_t$ such that $\col(u)=c_i$. Without loss of generality, let $u$ be in $N_{G_{t_1}}[v]$. Note that $u\in V_{t_1}\setminus X_t$. Assign $f_1(v)=W$ and $f_2(v)=R$. Thus $u$ is the unique vertex in $N_{G_{t_1}}$ such that $\col(u)=c_i$. Therefore $\col |_{V_{t_1}}$ is a $[t_1,c_1,\gamma_1,f_1]$ - good coloring and $d[t_1,c_1,\gamma_1,f_1]=true$. We will show that $d[t_2,c_2,\gamma_2,f_2]=true$. For contradiction, assume not. Therefore, in any coloring of $G_{t_2}$, $v$ is already conflict free dominated by a vertex with color $c_i$. Therefore, in the coloring $\col |_ {V_{t_2}} : V_{t_2} \rightarrow \{c_o,\dots,c_k\}$ there exists $w \in N_{G_{t_2}}[v]$ such that $\col(w)=c_i$. This implies that $G_t$ has two vertices $u, w \in N_{G_t}[v]$ such that $\col(u)=\col(w)=c_i$. This contradicts that $\col$ is a $(t,c,\gamma,f)$-good coloring. 
Similarly we can prove that $d[t,c,\gamma,f]=true$ when $d[t_1,c_1,\gamma_1,f_1]=d[t_2,c_2,\gamma_2,f_2]=true$.
\vspace{5pt}

\noindent\emph{Case 2: $f(v)=B$} : Assume $c(v)=\gamma(v)=c_i$. Clearly $\col(v)=c_i$. and there does not exist $u \in N_{G_t}(v)$ such that $\col(u)=c_i$. Therefore $\col |_ {V_{t_1}}$ and $\col |_ {V_{t_2}}$ respectively are $[t_1,c_1,\gamma_1,f_1]$ - good  and $[t_2,c_2,\gamma_2,f_2]$ - good colorings. Now assume $c(v) \ne \gamma(v)$. Then there exists $u \in N_{G_t}(v)$ such that $\col(u) =c_i$. Without loss of generality, assume $u \in V_{t_1}$. Assign $f_1(v)=B $ and $f_2(v)=C$. We can show that $d[t_1, c, \gamma, f_1]=d[t_2, c, \gamma, f_2]=true$ by similar arguments as in \emph{Case 1}.
\\

\noindent\emph{Case 3: $f(v)=C$ or $f(v)=W$} : When $f(v)=C$, $\col(v) \ne c_0$ and there does not exist $u \in N_{G_t}[v]$ such that $\col(u)=\gamma(v)$. Since $G_{t_1}$ and $G_{t_2}$ are subgraphs of $G_{t}$, $\col |_ {V_{t_1}}$ and $\col |_ {V_{t_2}}$ respectively are $[t_1,c_1,\gamma_1,f_1]$ - good  and $[t_2,c_2,\gamma_2,f_2]$ - good colorings. Similarly we can prove the reverse direction. The case $f(v)=R$ is similar.
\end{proof}

\begin{lemma}~\label{join_node_scfc}
Recurrence for the join node for \textsc{k-Strong Conflict-Free Coloring} is correct.
\end{lemma}
\begin{proof}

Consider the join node $X_t$. Let $D[t,c,\Gamma,f]=true$ and ${\sf col}: V_{t} \rightarrow \{c_0,c_1,\ldots, c_k\}$ be a $(t,c,\Gamma,f)$-good coloring. We prove that there exists $[t_1,c_1,\Gamma_1,f_1]$ and $[t_2,c_2,\Gamma_2,f_2]$, such that the pair is  $(t,c,\Gamma,f)$-consistent and $D[t_1,c_1,\Gamma_1,f_1] =true$ and $D[t_2,c_2,\Gamma_2,f_2] = true$.  For every vertex $v$, we assign $c_1(v)=c_2(v)=c(v)$. For a given $i \in \{1,\dots,k\}$, if $\Gamma(v)[i] \in \{0,\hat{1}\}$, then assign $\Gamma_1(v)[i] = \Gamma_2(v)[i]=\Gamma(v)[i]$. 
\vspace{5pt}

\noindent \emph{Case 1: $f(v)=W$} : 
We assign values for $\Gamma_1(v)$ and $\Gamma_2(v)$ as follows. For a given $i$, let  $\Gamma(v)[i] =1$. Therefore, there exists a unique neighbor $u$ of $v$ in $G_t$ such that $\col(u) =c_i$. Since $X_t$ induces an independent set in $G_t$, $u \in V_{t_1}\setminus X_t$ or $u \in V_{t_2}\setminus X_t$. Without loss of generality, let $u \in V_{t_1}\setminus X_t$. Assign $\Gamma_1(v)[i] =1$ and  $\Gamma_2(v)[i] =\hat{1}$. 

Now we assign $f_1(v)$(and similarly $f_2(v)$) as follows . If there exists an $i, 1 \leq i \leq k$ such that $\Gamma_1(v)[i] =1$, then assign $f_1(v)=W$, else assign $f_1(v)=R$.  By similar arguments as given in the proof of lemma~\ref{jointnodelemma1}, we can show that $\col |_ {V_{t_1}}$ and $\col |_ {V_{t_2}}$ respectively are $[t_1,c_1,\Gamma_1,f_1]$ - good  and $[t_2,c_2,\Gamma_2,f_2]$ - good colorings. 
\vspace{10pt}
\\
\noindent \emph{Case 2: $f(v)=B$} :
Assign $f_1(v)=f_2(v)=B$ and $\Gamma_1(v)[c(v)]=\Gamma_2(v)[c(v)]=1$. For $i$ such that  $c_i \neq c(v)$, assign $\Gamma_1$ and $\Gamma_2$ as follows. There exists a unique neighbor $u$ of $v$ in $G_t$ such that $\col(u) =c_i$. Since $X_t$ induces an independent set in $G_t$, $u \in V_{t_1}\setminus X_t$ or $u \in V_{t_2}\setminus X_t$. Without loss of generality, let $u \in V_{t_1}\setminus X_t$. Assign $\Gamma_1(v)[i] =1$ and  $\Gamma_2(v)[i] =\hat{1}$. $\col |_ {V_{t_1}}$ and $\col |_ {V_{t_2}}$ respectively are $[t_1,c_1,\Gamma_1,f_1]$ - good  and $[t_2,c_2,\Gamma_2,f_2]$ - good colorings.
\vspace{9pt}

\noindent \emph{Case 3: $f(v)=R$} :
Assign $f_1(v)=f_2(v)=R$. Since all $\Gamma \in \{0,\hat{1}\}$, the assignments $\Gamma_1$ and $\Gamma_2$ are covered before. In this case also, it is easy to see that $\col |_ {V_{t_1}}$ and $\col |_ {V_{t_2}}$ respectively are $[t_1,c_1,\Gamma_1,f_1]$ - good  and $[t_2,c_2,\Gamma_2,f_2]$ - good colorings.

\end{proof}